\documentclass[final]{cvpr}

\usepackage{times}
\usepackage{epsfig}
\usepackage{graphicx}
\usepackage{amsmath}
\usepackage{amssymb}
\usepackage{multirow}
\usepackage{subcaption}
\usepackage{anyfontsize}
\usepackage[svgnames,table]{xcolor}

\usepackage[pagebackref=true,breaklinks=true,colorlinks,bookmarks=false]{hyperref}

\def\cvprPaperID{5416}
\def\confYear{CVPR 2021}

\newcommand{\textapprox}{$_{\widetilde{~}}$}
\newcommand{\LL}{\scriptscriptstyle{LL\mid}}
\newcommand{\comma}{,\;}
\title{Efficient Multi-Stage Video Denoising with Recurrent Spatio-Temporal Fusion}

\author{
    Matteo Maggioni\thanks{Authors contributed equally.}, Yibin Huang\footnotemark[1], Cheng Li\footnotemark[1], Shuai Xiao, Zhongqian Fu, Fenglong Song\\
    {Huawei Noah's Ark Lab}\\
    {\tt\footnotesize \{matteo.maggioni, huangyibin1, licheng89, xiaoshuai7, fuzhongqian, songfenglong\}@huawei.com}
}

\begin{document}
    
    \maketitle
    
    \begin{abstract}
        In recent years, denoising methods based on deep learning have achieved unparalleled performance at the cost of large computational complexity. In this work, we propose an Efficient Multi-stage Video Denoising algorithm, called EMVD, to drastically reduce the complexity while maintaining or even improving the performance. First, a fusion stage reduces the noise through a recursive combination of all past frames in the video. Then, a denoising stage removes the noise in the fused frame. Finally, a refinement stage restores the missing high frequency in the denoised frame. All stages operate on a transform-domain representation obtained by learnable and invertible linear operators which simultaneously increase accuracy and decrease complexity of the model. A single loss on the final output is sufficient for successful convergence, hence making EMVD easy to train. Experiments on real raw data demonstrate that EMVD outperforms the state of the art when complexity is constrained, and even remains competitive against methods whose complexities are several orders of magnitude higher. Further, the low complexity and memory requirements of EMVD enable real-time video denoising on commercial SoC in mobile devices.
    \end{abstract}
    
    \section{Introduction}
    
    Even with the advance of technology, digital images are invariably affected by several inherent or external disturbing factors due to the stochastic nature of the image formation processes (e.g., photon counting)~\cite{foi2008noisemodel}, use of compact camera hardware (e.g., mobile sensors or lenses)~\cite{yue2020supervised}, and/or challenging acquisition settings (e.g., low light). Because of this, a number of methods (i.e., an image processing pipeline) must be applied to the low-quality observed data to generate a final high-quality output image. Denoising is particularly important because it is typically at the beginning of the pipeline, and thus its output has a direct effect on all other operations~\cite{wang2020practical}.
    
    In the past decades, a plethora of image denoising algorithms have been proposed in the literature~\cite{buades2005review,dabov2007bm3d}, but the current state of the art is dominated by deep learning methods based on convolutional neural networks (CNNs)~\cite{zhang2017dncnn,liu2018nlrn,liu2018wcnn}. Video denoising models exploits the temporal correlation inherently present in natural videos and thus achieve better performance than single-frame methods~\cite{maggioni2012vbm4d,arias2018vnlb,mildenhall2018kpn,tassano2020fastdvdnet,yue2020supervised}, however their computational requirements make real-time processing unattainable on most hardware unless some compromise in image quality is made~\cite{ehman2018realtime}.
    
    In this work we propose EMVD, an Efficient Multi-stage Video Denoising method to drastically reduce the complexity required to achieve high-quality results. Firstly, noise in the input frame is reduced by recursively fusing all past frames in the video. Then, a denoising stage removes any remaining noise in the fused image. Finally, a refinement stage is applied to the denoised image to further improve its quality by adaptively restoring the high-frequency details. All stages are performed within a domain generated by learnable and invertible linear transform operators that jointly decorrelate color and frequency information. As can be seen in Fig.~\ref{fig:first_example}, the proposed EMVD is able to outperform more complex state-of-the-art methods~\cite{yue2020supervised} at a fraction of the computational cost. In summary, the main contributions of this work are
    \begin{itemize}
        \item \textbf{High-Quality Efficient Denoising.} 
        The proposed EMVD leverages spatio-temporal correlation of natural videos through specialized processing stages, namely temporal fusion, spatial denoising, and spatio-temporal refinement. This design allows to significantly reduce the model complexity without compromising its denoising capabilities.
        \item \textbf{Interpretable Design.} 
        All stages in the proposed EMVD have a clear objective and will naturally converge to the desired behavior without any explicit supervision. As a result, the inner workings of the proposed EMVD can be easily inspected and controlled at both inference and training time.
        \item \textbf{Learnable and Invertible Transforms.} 
        Linear transform operators, implemented as learnable convolutional layers, are used to optimally decorrelate color and frequency information. The learnable parameters are regularized in the loss to ensure transform invertibility. This simultaneously allows to reduce complexity and increase accuracy of the model.
    \end{itemize}

    \section{Related works}
    \label{section:related}
    
    \textbf{Image Denoising.} Denoising is a long-studied research topic~\cite{buades2005review,dabov2007bm3d}, however the numerous works proposed in the recent past~\cite{zhang2017dncnn,liu2018nlrn,liu2018wcnn,chang2020spatial} suggest that the interest towards this problem is still very active, especially in the more challenging case of real raw data~\cite{brooks2019unprocessing,guo2019toward,kim2020transfer,chen2019motiondark}.

    Classical methods heavily exploit nonlocal image priors~\cite{buades2005nlm,dabov2007bm3d} and still provide outstanding performance today. Despite being designed for Gaussian noise, such methods can be also applied to real raw data when a variance-stabilizing transformation (VST) is used~\cite{makitalo2013vst}. More recently --and since the pioneering work~\cite{dong2014learning}-- deep learning and CNNs have become the \textit{de-facto} standard solution for all vision problems, including denoising. CNN-based methods most notably leverage residual learning~\cite{zhang2017dncnn}, wavelet decomposition~\cite{liu2018wcnn}, attention mechanisms~\cite{liu2018nlrn}, and spatially adaptive processing~\cite{hang2019pacnet,chang2020spatial}. 
    
    \textbf{Video Denoising.} Natural videos exhibit a strong correlation along the temporal dimension, i.e., pixels at corresponding locations in consecutive frames are likely to be very similar. One viable strategy to account for temporal correlation is to explicitly estimate the motion in the video with, e.g., block matching~\cite{maggioni2012vbm4d}, optical flow~\cite{caballero2017real,sajjadi2018frvsr}, kernel-prediction networks~\cite{mildenhall2018kpn}, or deformable convolutions~\cite{tian2020tdan,wang2019edvr}. With that, frames can be aligned before filtering to aid the restoration task. However, since motion estimation is a challenging and computationally demanding problem, a different line of research suggests to implicitly deal with motion through, e.g., spatio-temporal attention modules which recursively aggregate features at different time steps~\cite{jo2018deep,yi2019progressive,wang2019edvr,davy2019vnl}. These methods can be further categorized into multi-frame approaches, whose inputs include several consecutive frames that are jointly processed by the model~\cite{maggioni2012vbm4d,claus2019videnn,tassano2019dvdnet,tassano2020fastdvdnet,yue2020supervised}, or recurrent approaches where images~\cite{sajjadi2018frvsr,ehman2018realtime} and/or features~\cite{godart2018deep,fuoli2019rlsp} obtained from the previous time step are used as additional input to estimate the current frame. Several models have been designed with efficiency in mind, yet real-time computation is still unattainable on most hardware~\cite{fuoli2019rlsp,tassano2020fastdvdnet}, or performance is not on-par with the state of the art~\cite{ehman2018realtime}.
    
    \begin{figure*}[ht!]
        \centering
        \includegraphics[trim=0cm 7.5cm 18.75cm 0cm,clip,width=0.8\textwidth]{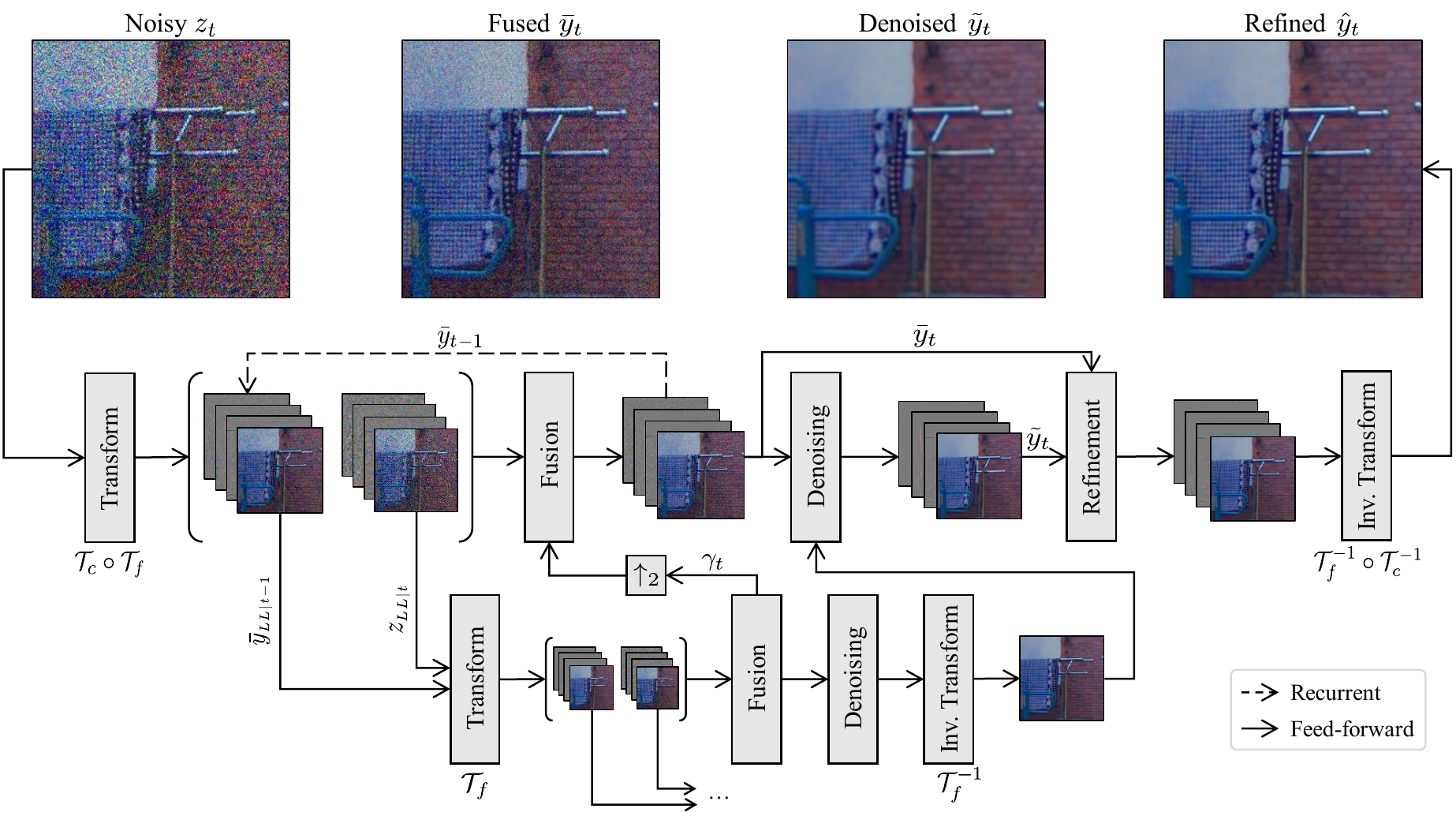}
        \caption{Architecture of the proposed multi-stage video denoising EMVD. Refer to Section~\ref{section:method} for details.}
        \label{fig:architecture} 
    \end{figure*}
    
    \section{Method}
    \label{section:method}

    \subsection{Observation Model}
    
    The goal of our denoising algorithm is to obtain an estimate of the clean video from the observed noisy data at a very low complexity. The observation model is defined as
    \begin{equation}
        z_t(x) = y_t(x) + \eta_t(x),
        \label{eq:raw_noise}
    \end{equation}
    where $t \in T \subset \mathbb{N}$ is the temporal index of the frame in the video, $x \in X \subset \mathbb{N}^2$ is a spatial pixel position in the frame, $z \in \mathbb{R}^{H \times W \times C}$ is the observed noisy raw video in packed form~\cite{gharbi2016jdd} having $H \times W$ resolution and $C=4$ color channels (e.g., $RG_1G_2B$), $y$ is the underlying (unknown) noise-free data to be estimated, and $\eta_t \sim \mathcal{N}\left(0, \sigma^2_t(y_t) \right)$ is a noise realization drawn from a heteroskedastic Gaussian distribution with signal-dependent variance
    \begin{equation}
        \sigma^2_t(y_t) = a_t y_t + b_t
        \label{eq:raw_variance}
    \end{equation}
    modeling signal-dependent (shot) and signal-independent (read) noise sources parametrized by $a_t$ and $b_t \in \mathbb{R}$, respectively. Since many robust estimation methods exist~\cite{foi2008noisemodel, azzari2014noiseestimation}, we assume such noise parameters to be known for the given sensor and camera ISO.
    
    \subsection{Learnable Invertible Transforms}
    
    The proposed method employs learnable transform operators --inspired by YUV and wavelet transforms-- to decorrelate color and frequency information of the raw data. These operators are linear and designed to be invertible, and thus can be implemented as standard convolutional operations with deliberately regularized weights.
    
    \textbf{Color Transform.} The color transform $\mathcal{T}_c$ is implemented as a point-wise convolution whose kernel is constrained to be an orthonormal matrix $M \in \mathbb{R}^{C \times C}$ which decorrelates the $C=4$ colors (e.g., $RG_1G_2B$) in the CFA image to a luminance-chrominance representation~\cite{buades2020cfa}. Being the color transform orthonormal, its inverse $\mathcal{T}^{-1}_c$ is simply implemented as another point-wise convolution with kernel initialized as $M^\prime = M^\top$. In this work, the weights in the forward and inverse matrices are not shared to allow more degrees of freedom. The resulting transform is therefore biorthogonal and its invertibility is enforced by a loss term defined as
    \begin{equation}
        \mathcal{L}_c = \big\lvert\big\lvert M \cdot M^\prime - I_C \big\rvert\big\rvert_F^2,
        \label{eq:color_loss}
    \end{equation}
    where $\cdot$ is matrix multiplication, and $I_C$ is the identity matrix of rank $C$, and $\parallel\cdot\parallel_F$ denotes the Frobenius norm.
    
    \textbf{Frequency Transform.} Inspired by biorthogonal wavelets, we design a transform $\mathcal{T}_f : \mathbb{R}^{H \times W \times C} \rightarrow \mathbb{R}^{H/2 \times W/2 \times 4C}$ to decorrelate the input frequencies into four half-resolution components, namely the low-pass $LL$ and high-pass $\{LH, HL, HH\}$ subbands. The transform is again linear and thus can be implemented as a strided convolution with four $n \times n$ kernels initialized as the outer product of some chosen wavelet decomposition filters $\psi \in \mathbb{R}^{2 \times n}$, being $n \in \mathbb{N}^+$ the (even) length of the wavelet filters (e.g., $n=2$ for Haar). Conversely, the inverse operator $\mathcal{T}^{-1}_f$ is implemented as a transposed convolution with kernels initialized this time from the corresponding reconstruction filters $\phi \in \mathbb{R}^{2 \times n}$. Note that the same $\mathcal{T}_f$ can be recursively applied on the $LL$ subband to produce a multi-scale decomposition of the input.

    We enforce invertibility of the transform by adding a loss term on the matrix form of the filters as
    \begin{equation}
        \mathcal{L}_f = \big\lvert\big\lvert \psi \cdot \phi^\top - I_2 \big\rvert\big\rvert_F^2,
        \label{eq:frequency_loss}
    \end{equation}
    where $I_2$ is the identity matrix of rank $2$. As such, the proposed method is learning the 1-D filter representation of the frequency transform, and not the convolutional form of the kernel. A different strategy would be forcing the learned filters to follow a wavelet parametric model~\cite{wolter2020wavelet}. Thus our approach might generate filters that do not satisfy basic wavelet properties, however our filters have more degrees of freedom and still produce an invertible transformation.
    
    Note that the composition of the color and frequency transform $\mathcal{T}_c \circ \mathcal{T}_f$ is still linear and invertible. A joint application of the two transforms allows to simultaneously increase accuracy and reduce complexity of the model because the energy of the meaningful part of the image is decorrelated from the energy of the noise and, at the same time, spatial resolution of the data is halved. A diagram of the proposed EMVD is illustrated in Fig.~\ref{fig:architecture}. In the remainder of this paper, we will assume the data to be given already in the transform domain, thus, for the sake of notation simplicity, we will omit the transform operators.
    
    \subsection{Fusion Stage}
    \label{section:fusion}
    
    The first processing stage of the proposed EMVD is temporal fusion. The objective of this stage is to maximally reduce the noise present in the image using the temporal correlation inherently present in the video without introducing any temporal artifact or degrading image structures. Formally, fusion is defined as a recursive convex combination
    \begin{equation}
        \bar{y}_t(x) = \bar{y}_{t-1}(x) \bar{\gamma}_{t-1}(x) + z_t(x) \gamma_t(x),
        \label{eq:fusion_equation}
    \end{equation}
    where $z_t$ is the transformed noisy frame at time $t$, $\bar{y}_{t-1}$ is the transformed fused output frame at previous time $t-1$, and $\gamma \in \mathbb{R}^{H/2 \times W/2 \times 1}$ are non-negative convex weights satisfying the condition $\bar{\gamma}_{t-1}(x) + \gamma_t(x) = 1$ at any given transform-domain location $x \in \chi \subset \mathbb{N}^{H/2 \times W/2 \times 4C}$. The initial condition for \eqref{eq:fusion_equation} is $\bar{y}_{0} \equiv z_0$, as no previous frame is available at time $t = 0$. Note that the number of channels of $\gamma$ is $1$, so fusion of the full $4C$ input channels is achieved by element-wise broadcasting. As a result, we apply the same weights to each channel (i.e., subband) of the input frames. Different weights could be predicted for different input channels, but this would significantly increase the difficulty of the prediction task, as well as the memory requirements. 

    \begin{figure}[t]
        \centering
        \includegraphics[trim=0cm 11.7cm 24.0cm 0cm,clip,width=0.95\columnwidth]{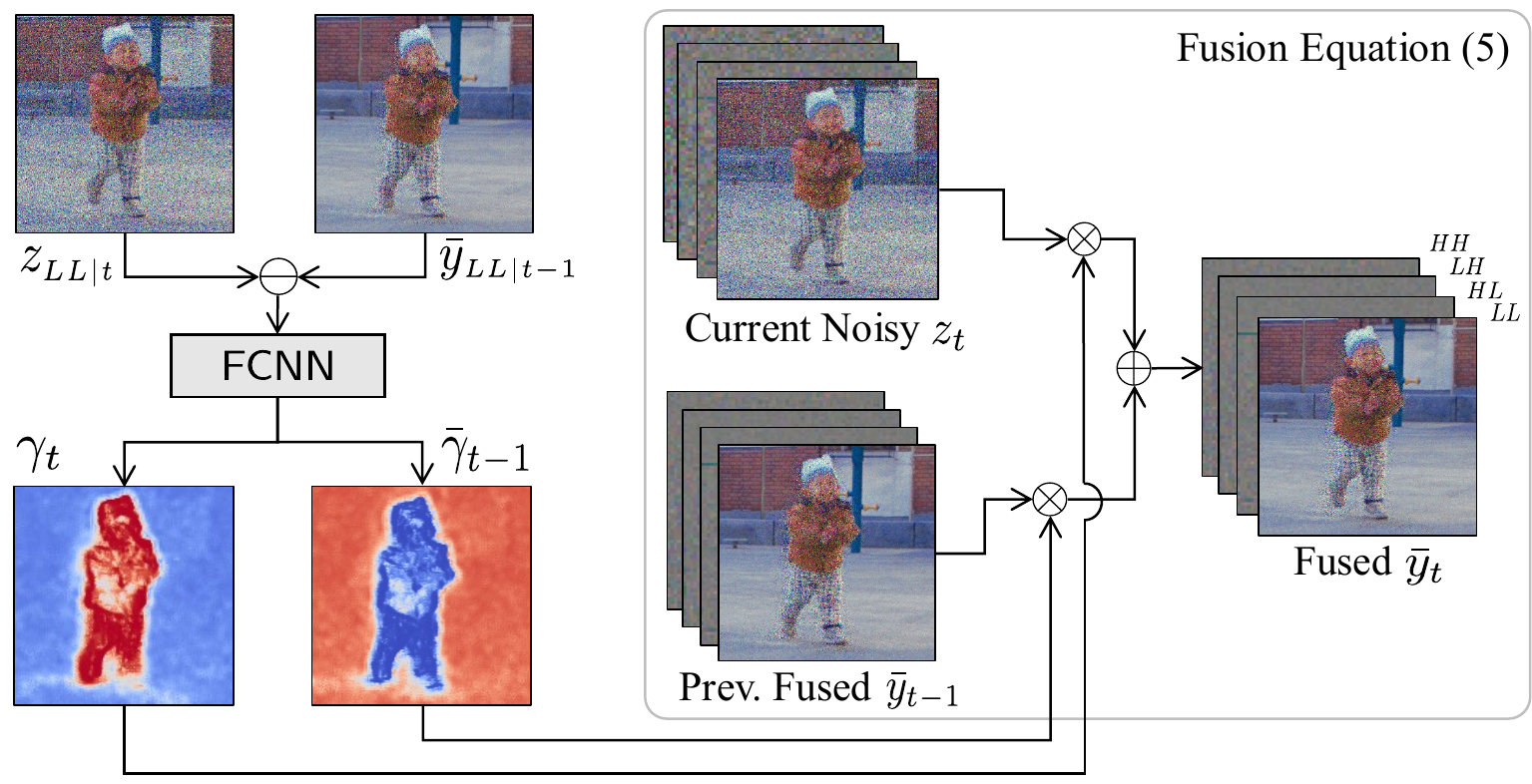}
        \caption{The predicted weights $\gamma_t$ discriminate the dynamic (red) and static (blue) regions between consecutive frames to minimize the noise in the output fused image.}
        \label{fig:fusion} 
    \end{figure}
    
    The weights in \eqref{eq:fusion_equation} are predicted by a fusion network, which we call $\mathsf{FCNN}$, defined as follows:
    \begin{equation}
        \big\{\gamma_t, \bar{\gamma}_{t-1}\big\} = \mathsf{FCNN}\Big( \left\lvert z_{\LL t} - \bar{y}_{\LL t-1}\right\rvert \comma \hat{\sigma}^2_t \Big),
        \label{eq:fcnn}
    \end{equation}
    where $\lvert\cdot\rvert$ denotes absolute value, $z_{\LL t}$ is the low-pass of the transformed noisy input frame, $\bar{y}_{\LL t-1}$ is the low-pass of the previous fused frame, and $\hat{\sigma}^2_t = \sigma^2_t\left(z_{\LL t}\right)$ is the variance of the input frame computed as in \eqref{eq:raw_variance}. Note that the variance is approximated using the low-pass of $z_t$ as proxy for the (unknown) noise-free $y_t$. In order to maintain convexity of \eqref{eq:fusion_equation}, the output layer of $\mathsf{FCNN}$ is activated by a sigmoid function. Fig.~\ref{fig:fusion} illustrates the diagram of the fusion network; note how the predicted weights $\gamma_t$ clearly separate the dynamic regions from the static ones. As a result, the output fused image is generated by adaptively averaging the incoming frames. Note that fusion is performed at a lower image resolutions, a single position at any given scale $i$ actually corresponds to a $2^i\times 2^i$ neighborhood in the original image, hence allowing some degree of motion compensation. As shown in Fig.~\ref{fig:architecture}, when multiple scales are available, the weights obtained from the fusion stage at the lower scale are upsampled and concatenated to the input of \eqref{eq:fcnn} to provide additional guidance information.

    Finally, \eqref{eq:fcnn} can be interpreted as a special case of kernel-predicting network~\cite{mildenhall2018kpn} with $1 \times 1$ kernels, thus \eqref{eq:fusion_equation} can be trivially extended to a general (convolutional) form as
    \begin{equation}
        \bar{y}_t(x) = \bar{y}_{t-1}(x) \circledast \bar{k}_{t-1}(x) + z_t(x) \circledast k(x),
        \label{eq:conv_fusion_equation}
    \end{equation}
    where $\circledast$ denotes convolution, $k$ is the spatially adaptive kernel of size $p \times p$ applied to the noisy frame, and $\bar{k}_{t-1}$ is the kernel of size $\bar{p} \times \bar{p}$ applied to the previous one. Practically, the kernels in~\eqref{eq:conv_fusion_equation} can be obtained by letting the output layer of the fusion network predict $\bar{p}^2+p^2$ channels activated by a softmax function to ensure that the fusion equation is still convex.

    \subsection{Denoising Stage}
    
    Noise in the output fused image $\bar{y}_t$ is reduced by \eqref{eq:fusion_equation} but not completely. This inevitably occurs when, e.g., motion cannot be compensated effectively or when the amount of processed temporal data is not enough to increase the signal-to-noise ratio (SNR) in the frame. Thus, we use a denoising network, called $\mathsf{DCNN}$, to remove any remaining noise in $\bar{y}_t$ as
    \begin{equation}
        \tilde{y}_t = \mathsf{DCNN}\Big( \bar{y}_t \comma z_{\LL t} \comma \bar{\sigma}^2_t \Big),
        \label{eq:dcnn}
    \end{equation}
    where $\tilde{y}_t$ is the denoised image, and $\bar{\sigma}^2_t$ is the noise variance of the fused image $\bar{y}_t$. The input also includes the low-pass of the current noisy frame $z_{\LL t}$ so that the network has the chance to extract valuable information from the unadulterated noisy input. When multiple scales are available, the image estimated at the lower scale is concatenated to the input of \eqref{eq:dcnn}.
    
    Denoising the fused image $\bar{y}_t$ is easier than directly denoising the input $z_t$, but the form of the variance $\bar{\sigma}^2_t$ is highly complex as it depends on the signal-dependent variance at frame $t$ as well as on the cumulative effect fusing all previous frames $t \in \{0, \ldots, t-1\}$. Nevertheless, fusion itself is linear, so we are able to define a recursive formulation of the fused variance using basic statistical properties\footnote{$\mathbb{V}\text{ar}[aX+bY] = a^2\mathbb{V}\text{ar}[X] + b^2\mathbb{V}\text{ar}[Y] + 2ab\mathbb{C}\text{ov}[X,Y]$} by expanding \eqref{eq:fusion_equation} into \eqref{eq:raw_variance} as
    \begin{align}
        \bar{\sigma}^2_t &\equiv \sigma^2_t\left(\bar{y}_{\LL t}\right) \nonumber \\
        &= \bar{\gamma}^2_{t-1} \sigma^2_{t-1}\left(\bar{y}_{\LL t-1}\right) + \gamma^2_t \sigma^2_t\left(z_{\LL t}\right),
        \label{eq:fusion_raw_variance}
    \end{align}
    where the initial condition is $\sigma^2_t\left(\bar{y}_{\LL 0}\right) \equiv \sigma^2_t\left(z_{\LL 0}\right)$, and the covariance term is zero as we assume that the noise is temporally independent. Note that, since $\gamma_t(x) \leq 1$ for all $x$ and $t$, the variance $\eqref{eq:fusion_raw_variance}$ is (non-strictly) decreasing with time, thereby justifying the intuitive idea that fusion will progressively reduce the noise in the input.

    \subsection{Refinement Stage}

    \begin{figure}[t]
        \centering
        \includegraphics[trim=0cm 2.5cm 14cm 0cm,clip,width=\columnwidth]{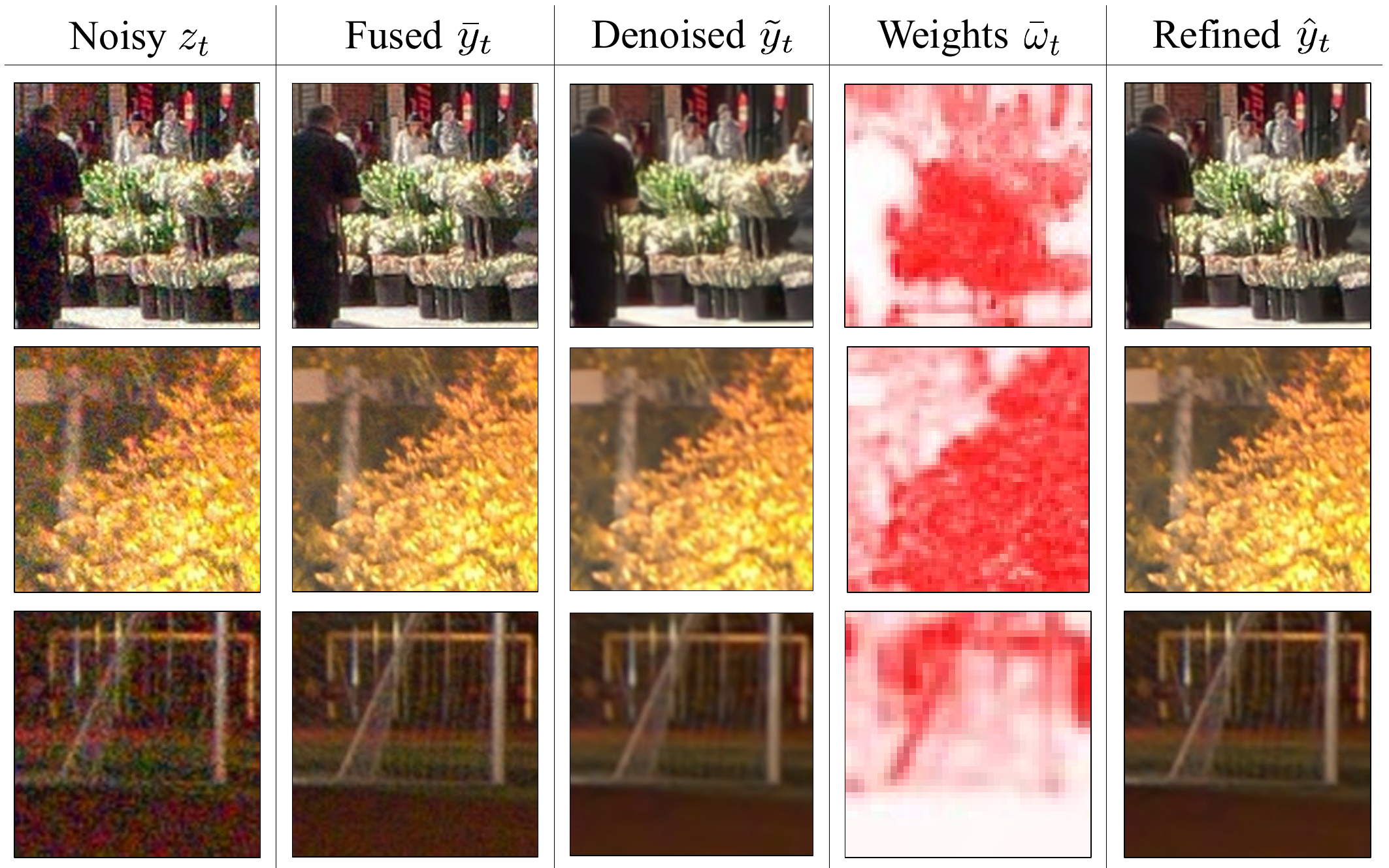}
        \caption{The red regions in the weights $\omega_t$ identify the high-frequency (e.g., edges and textures) information of the fused image used to refine the denoised image.}
        \label{fig:refinement} 
    \end{figure}
    
    Any denoising method is likely to introduce artifacts and loss of image details, especially when SNR of the input image is poor or when the complexity of model is significantly constrained. Thus, we propose to use a final refinement stage to combine the detailed --but still noisy-- fused image $\bar{y}_{t}$ with the noise-free --but likely oversmoothed-- denoised image $\tilde{y}_t$. In doing so, we expect to restore the fine details and textures potentially removed by the denoising, a task which is facilitated by our learned frequency representation which naturally decorrelates low- to high-frequency information. Formally, we seek to solve a refinement equation
    \begin{equation}
        \hat{y}_t(x) = \bar{y}_{t}(x) \bar{\omega}_{t}(x) + \tilde{y}_t(x) \tilde{\omega}_t(x),
        \label{eq:refinement_equation}
    \end{equation}
    with convex weights satisfying $\bar{\omega}_{t}(x) + \tilde{\omega}_{t}(x) = 1$ for all $x \in \chi$. The refinement weights are predicted by yet another network, called $\mathsf{RCNN}$, operating as
    \begin{equation}
        \big\{\tilde{\omega}_t(x), \bar{\omega}_{t}\big\} = \mathsf{RCNN}\Big( \tilde{y}_{t} \comma \bar{y}_{t} \comma \bar{\sigma}^2_t \Big),
        \label{eq:rcnn}
    \end{equation}
    where convexity of the output weights is again ensured by applying a sigmoid activation on the final output layer. Although the formulation~\eqref{eq:refinement_equation} is equivalent to the fusion equation~\eqref{eq:fusion_equation}, the refinement weights $\omega$ have a markedly different meaning than the fusion weights $\gamma$. In fact, as shown in Fig.~\ref{fig:refinement}, refinement weights are used to identify high-frequency information from the fused image whereas fusion weights are used to recursively aggregate consistent temporal information across consecutive frames to reduce the effect of the noise. Interestingly, no explicit supervision is required to converge to this behavior.

    As shown in Fig.~\ref{fig:architecture}, the refinement network is only used at the higher scale of the frequency decomposition, even when lower scales are available. Finally, we highlight that~\eqref{eq:rcnn} can be extended to predict kernels of arbitrary size analogously to~\cite{mildenhall2018kpn}.

    \section{Experiments}
    \label{section:experiments}

    We compare the proposed EMVD against various state-of-the-art video denoising methods, namely VBM4D~\cite{maggioni2012vbm4d},  RViDeNet~\cite{yue2020supervised}, FastDVDnet~\cite{tassano2020fastdvdnet}, and EDVR~\cite{wang2019edvr}. In our experiments, we show performance of the aforementioned models at various levels of complexity, measured as floating-point operations (FLOPs). Additional technical details and experiments can be found in the supplementary materials.  
    
    To evaluate our model, we use the video benchmark dataset proposed in~\cite{yue2020supervised}. This consists of a real raw video dataset (CRVD) captured by a SONY IMX385 sensor and a synthesized dataset (SRVD) generated from~\cite{chen2018sid}; all videos have five different ISO levels ranging from 1600 to 25600. Following~\cite{yue2020supervised}, we use SRVD videos plus the scenes 1--6 from CRVD for training, hence keeping CRVD scenes 7--11 for objective validation. CRVD also includes few outdoor noisy raw videos without ground-truth which we use as test set to subjectively asses visual quality. 

    \begin{table*}[ht!]
        \centering
        \small
        \setlength{\tabcolsep}{2pt}
        \begin{subtable}{\columnwidth}
            \centering          
            \begin{tabular}{ccccccccc}
                \hline
                GFLOPs&   Recurrent    & $z_{\LL t}$  & $\mathcal{T}_c$ & $\mathcal{T}_f$  & $\mathsf{RCNN}$ &$\sigma^2$& PSNR / SSIM \\
                \hline      \hline  
                \rowcolor{LightYellow}
                5.38& $ \bar{y}_{t-1}   $ & $\checkmark$  & $\checkmark$ & $\checkmark$ & $\checkmark$  & $\checkmark$ & \textbf{42.63} / \textbf{0.9851 } \\
                5.38& $ \hat{y}_{t-1}   $ & $\checkmark$  & $\checkmark$ & $\checkmark$ & $\checkmark$  & $\checkmark$ & 42.38 / 0.9840 \\
                
                5.12& $ \bar{y}_{t-1}   $ & $\times$      & $\checkmark$ & $\checkmark$ & $\checkmark$  & $\checkmark$ & 41.87 / 0.9831 \\
                5.31& $ \bar{y}_{t-1}   $ & $\checkmark$  & $\times$     & $\checkmark$ & $\checkmark$  & $\checkmark$ & 42.36 / 0.9839 \\
                5.38& $ \bar{y}_{t-1}   $ & $\checkmark$  & $\checkmark$ & $\times$     & $\checkmark$  & $\checkmark$ & 42.35 / 0.9838 \\
                5.94& $ \bar{y}_{t-1}   $ & $\checkmark$  & $\checkmark$ & $\checkmark$ & $\times$      & $\checkmark$ & 42.46 / 0.9848 \\
                
                5.18& $ \bar{y}_{t-1}   $ & $\checkmark$  & $\checkmark$ & $\checkmark$ & $\checkmark$  & $\times$     & 41.39 / 0.9795 \\
                5.42& $ \hat{y}_{t-1}   $ & $\times$      & $\times $    & $\times$     & $\times$      & $\checkmark$ & 41.35 / 0.9737 \\
                \hline              
            \end{tabular}
            \caption{Network structure.}
            \label{tab:structure}
        \end{subtable}
        \begin{subtable}{\columnwidth}
            \centering
            \setlength{\tabcolsep}{6pt}
            \begin{tabular}{ccccc}
                \hline
                GFLOPs         & $\mathsf{FCNN}$ & $\mathsf{DCNN}$ & $\mathsf{RCNN}$ & PSNR / SSIM                       \\
                \hline  \hline
                2542.86        & 4 / 64          & 4 / 512         & 4 / 64          & \textbf{44.51} / \textbf{0.9897}  \\
                1105.65        & 4 / 64          & 6 / 256         & 4 / 64          & 44.48 / 0.9895                    \\
                \hline
                86.23          & 3 / 64          & 3 / 64          & 3 / 64          & 43.57 / 0.9881                    \\
                82.06          & 4 / 16          & 5 / 64          & 3 / 64          & 43.83 / 0.9883                    \\
                79.52          & 4 / 16          & 6 / 64          & 1 / 32          & 44.05 / 0.9890                    \\
                25.31          & 4 / 16          & 5 / 32          & 3 / 32          & 43.19 / 0.9869                    \\
                \hline
                9.85           & 4 / 16          & 5 / 16          & 3 / 16          & 42.73 / 0.9854                    \\
                \rowcolor{LightYellow}
                5.38           & 2 / 16          & 2 / 16          & 2 / 16          & 42.63 / 0.9851                    \\
                \hline
            \end{tabular}
            \caption{Number of convolutions / number of filters.}
            \label{tab:distribution}
        \end{subtable}
        \caption{Ablation study of the proposed EMVD evaluated on the raw CRVD dataset~\cite{yue2020supervised}.}
        \label{tab:ablation}
    \end{table*}
    
    \subsection{Training}
    \label{section:training}
    
    We use training sequences composed of $n$ patches of size $128 \times 128$ cropped at random spatio-temporal locations of the training videos, minding to preserve the CFA Bayer pattern~\cite{liu2019bayer}. Specifically, we use $n=3$ for RViDeNet, $n=5$ for FastDVDnet and EDVR, and $n=25$ for EMVD. Note that EMVD is a recurrent models, and thus benefit from large values of $n$ since the models are temporally unrolled during training to allow backpropagation through time. Differently, RViDeNet, FastDVDnet, and EDVR are multi-frame methods and thus $n$ is equal to the number of frames used in their inputs.
    
    The loss is defined as $\mathcal{L} = \mathcal{L}_r + \mathcal{L}_c + \mathcal{L}_f$, where $\mathcal{L}_r = \frac{1}{n} \sum_{t=1}^n \parallel \hat{y}_t - y_t \parallel_1$ denotes the mean L1 norm of the difference between each predicted $\hat{y}_t$ and ground-truth $y_t$ frame in the sequence, and $\mathcal{L}_c$ and $\mathcal{L}_f$ are terms constraining invertibility of the color~\eqref{eq:color_loss} and frequency~\eqref{eq:frequency_loss} transforms, respectively. We train the networks using Adam optimizer~\cite{kingma2014adam} with batch size 16 and initial learning rate 1e-4. We apply a piece-wise constant decay which reduces the learning rate by a factor of 10 every 100000 iterations. All models are trained for an initial 300000 iterations on the CRVD and SRVD dataset, and then fine-tuned for an additional 300000 iterations on CRVD only. We implemented the proposed EMVD with Huawei MindSpore~\cite{mindspore} and TensorFlow; both implementations show comparable accuracy and efficiency.

    For RViDeNet we directly utilize the model and weights provided by the authors, and we train versions with reduced complexity using the same three-stage procedure suggested in the original paper~\cite{yue2020supervised}. For VBM4D we perform a grid search on the target sRGB image to find the optimal parameters that maximize the validation PSNR.

    \subsection{Ablation}
    \label{section:ablation}
    
    \textbf{Baseline.} There are three CNNs involved in EMVD. Any backbone could be used, but, for the sake of efficiency, in all cases we use two convolutional layers ($3\times3$ kernels, 16 filters) followed by ReLU activation plus one final output convolution. The inputs (i.e., images and variance) are always concatenated before processing. The output layers of both fusion and refinement CNNs are activated by a sigmoid, but no activation is applied to the denoising CNN. We use three decomposition scales. The frequency transform is initialized with Haar kernels, and the color transform is initialized as in~\cite{buades2020cfa}. This configuration is highlighted in yellow in all tables and correspond to 5.38 GFLOPs.
    
    \textbf{Network Structure.} Table~\ref{tab:structure} reports an ablation study on the structure of proposed method. In the fusion~\eqref{eq:fcnn}, if we use the previous final output $\hat{y}$ instead of the previous fused image $\bar{y}$ PSNR decreases by 0.25dB; instead if we remove the low-pass noisy input $z_{\LL t}$ from the denoising~\eqref{eq:dcnn} PSNR decreases by 0.76dB. The color transform, while only costing 0.08 GFLOPs, provides a sizable 0.27dB boost, and if we replace the learnable frequency transform with pixel shuffling~\cite{shu2016subpx} the PSNR reduces by 0.28dB. Then, if we remove the refinement stage, while increasing capacity of the denoising network as not to change overall complexity, we observe a 0.17dB decrease. Next we show that removing the variance $\sigma^2$ input from all networks (i.e., a blind formulation) result in a very significant 1.24dB PSNR drop. Finally, in the last row, we show that the drop is even higher (1.28dB) when both fusion and refinement are disabled.
    
    \textbf{Capacity Distribution.} In Table~\ref{tab:distribution} we report how EMVD is affected by the number of filters and convolutions in each stage. Our experiments indicate that it is beneficial to dedicate more capacity to the denoising CNN (as denoising is the most difficult stage), while the capacity of both fusion and refinement CNNs can be reduced without significantly impacting performance.

    \textbf{Fusion Recurrence.} We use the previous fused image $\bar{y}_{t-1}$ as input of the fusion is preferable than using the previous output frame $\hat{y}_{t-1}$ or noisy frame $z_{t-1}$ as we aim to maximally reducing the noise without removing image details. In fact, using the noisy $z_{t-1}$ will at most decrease the noise variance by a factor of $2$, effectively reducing our recurrent model to a multi-frame one. Then, as reported in Table~\ref{tab:structure}, using the output $\bar{y}_{t-1}$ is also suboptimal. This might be counter-intuitive, but since denoising is applied to $\bar{y}_{t-1}$, then the recurrent variance~\eqref{eq:fusion_raw_variance} would no longer admit a closed-form definition (because denoising is nonlinear), and also some of the high-frequency information in the image might be oversmoothed or even missing.

    \begin{table}[t]
        \centering
        \small
        \vspace{-0.2cm}
        \begin{tabular}{cccc}
            \hline
            GFLOPs  & $\bar{p} \times \bar{p}$ & $p \times p$ & PSNR / SSIM   \\
            \hline \hline
            2542.86 & $1\times1$ & $1\times1$ & 44.51 / 0.9897   \\
            2544.25 & $3\times3$ & $1\times1$ & \textbf{44.58} / \textbf{0.9899}   \\
            2545.49 & $3\times3$ & $3\times3$ & \textbf{44.58} / \textbf{0.9899}   \\
            2546.73 & $5\times5$ & $1\times1$ & 44.48 / 0.9897   \\
            2550.44 & $5\times5$ & $5\times5$ & 44.51 / 0.9897   \\
            \hline
        \end{tabular}
        \vspace{-0.1cm}
        \caption{Ablation on the fusion kernel sizes of the proposed EMVD evaluated on the raw CRVD dataset~\cite{yue2020supervised}.}
        \label{tab:kpn_fusion}
        \vspace{-0.3cm}
    \end{table}

    \textbf{Fusion Prediction.} In Table~\ref{tab:kpn_fusion}, we compare validate kernel-predicting fusion~\eqref{eq:conv_fusion_equation} using various kernel sizes. The first row correspond to the top-performing baseline model defined in Table~\ref{tab:distribution} with element-wise fusion (i.e., $1\times1$ kernels). We compare different sizes (up to $5\times5$) for the kernels $\bar{p} \times \bar{p}$ and $p \times p$ applied to the previous fused image and to the current noisy image, respectively. We observe that it is more beneficial to use large kernels on the previous image, which is in fact where motion compensation is needed, and that this strategy improves PSNR by \textapprox0.1dB PSNR at a relatively limited increase in complexity

    \begin{figure*}[ht!]
   	\begin{subfigure}[t]{\textwidth}
   		\setlength{\tabcolsep}{0.75pt}
   		\centering
   		\begin{tabular}[b]{rcccccc}
   			{\footnotesize Model} & {\footnotesize VBM4D} & {\footnotesize FastDVDnet} & 
   			{\footnotesize EDVR} & {\footnotesize RViDeNet} & {\footnotesize Ours}& {\footnotesize Noisy} \\[-0.1cm]   			
   			{\scriptsize GFLOPs} & {\scriptsize -} &  {\scriptsize 664.99} & 
   			{\scriptsize 3088.98} & {\scriptsize 1965.04} & {\scriptsize 5.38} & \\
   			
   			\includegraphics[height=0.23\textwidth]{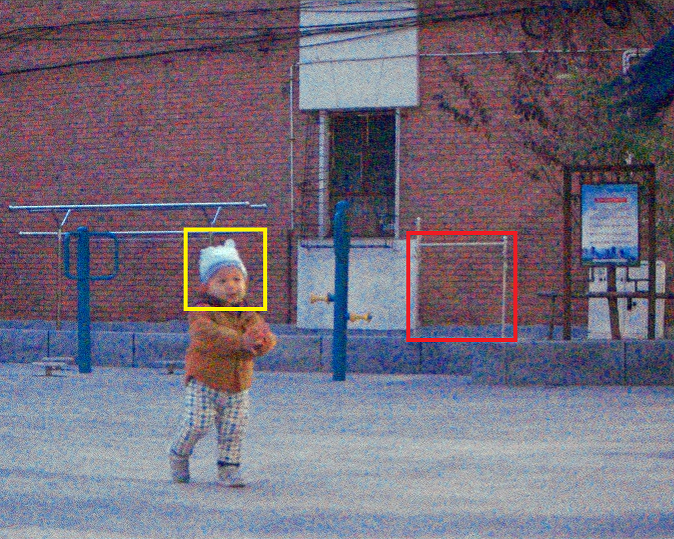} & 
   			\begin{tabular}[b]{c}
   				\includegraphics[width=0.112\textwidth]{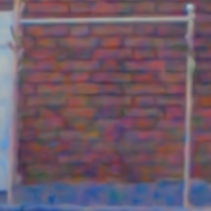} \\[-0.025cm]
   				\includegraphics[width=0.112\textwidth]{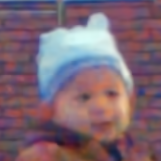}
   			\end{tabular} &
   			\begin{tabular}[b]{c}
   				\includegraphics[width=0.112\textwidth]{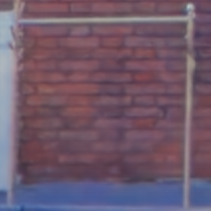} \\[-0.025cm]
   				\includegraphics[width=0.112\textwidth]{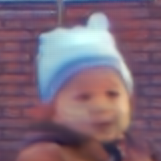}
   			\end{tabular} &
   			\begin{tabular}[b]{c}
   				\includegraphics[width=0.112\textwidth]{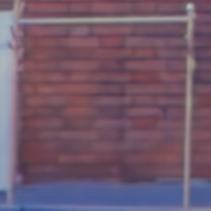} \\[-0.025cm]
   				\includegraphics[width=0.112\textwidth]{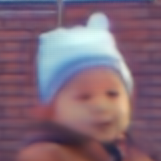}
   			\end{tabular} &
   			\begin{tabular}[b]{c}
   				\includegraphics[width=0.112\textwidth]{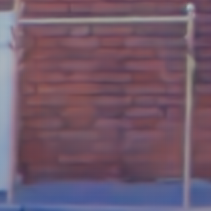} \\[-0.025cm]
   				\includegraphics[width=0.112\textwidth]{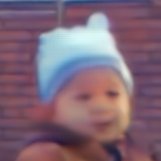}
   			\end{tabular} &
   			\begin{tabular}[b]{c}
   				\includegraphics[width=0.112\textwidth]{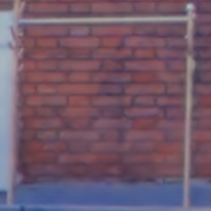} \\[-0.025cm]
   				\includegraphics[width=0.112\textwidth]{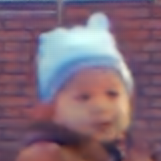}
   			\end{tabular} &
   			\begin{tabular}[b]{c}
   				\includegraphics[width=0.112\textwidth]{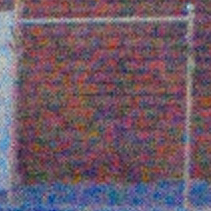} \\[-0.025cm]
   				\includegraphics[width=0.112\textwidth]{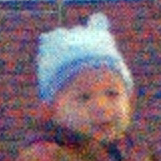}
   			\end{tabular}
   		\end{tabular}
        \vspace{-0.55cm}
   		\caption{Scene 2 -- ISO 25600}
        \vspace{0.1cm}
   	\end{subfigure}
   	\begin{subfigure}[t]{\textwidth}
   		\setlength{\tabcolsep}{0.75pt}
   		\centering
   		\begin{tabular}[b]{rccccccc}
   			\includegraphics[height=0.23\textwidth]{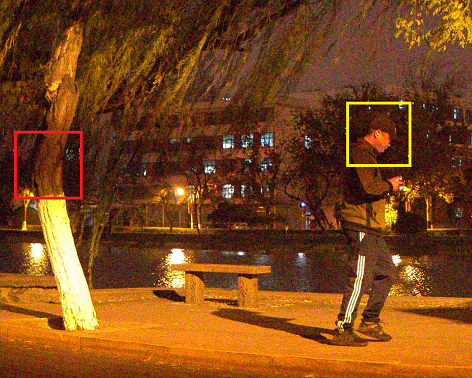} & 
   			\begin{tabular}[b]{c}
   				\includegraphics[width=0.112\textwidth]{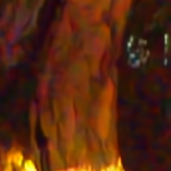} \\[-0.025cm]
   				\includegraphics[width=0.112\textwidth]{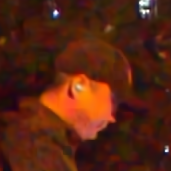}
   			\end{tabular} &
   			\begin{tabular}[b]{c}
   				\includegraphics[width=0.112\textwidth]{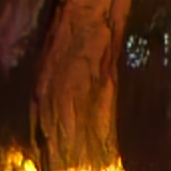} \\[-0.025cm]
   				\includegraphics[width=0.112\textwidth]{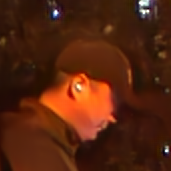}
   			\end{tabular} &
   			\begin{tabular}[b]{c}
   				\includegraphics[width=0.112\textwidth]{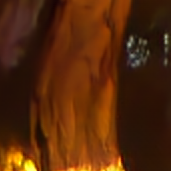} \\[-0.025cm]
   				\includegraphics[width=0.112\textwidth]{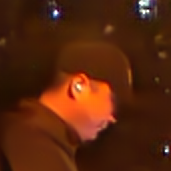}
   			\end{tabular} &
   			\begin{tabular}[b]{c}
   				\includegraphics[width=0.112\textwidth]{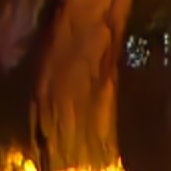} \\[-0.025cm]
   				\includegraphics[width=0.112\textwidth]{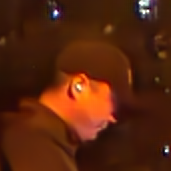}
   			\end{tabular} &
   			\begin{tabular}[b]{c}
   				\includegraphics[width=0.112\textwidth]{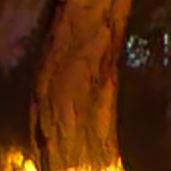} \\[-0.025cm]
   				\includegraphics[width=0.112\textwidth]{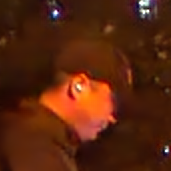}
   			\end{tabular} &
   			\begin{tabular}[b]{c}
   				\includegraphics[width=0.112\textwidth]{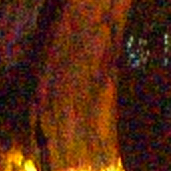} \\[-0.025cm]
   				\includegraphics[width=0.112\textwidth]{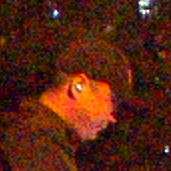}
   			\end{tabular}
   		\end{tabular}
        \vspace{-0.55cm}
   		\caption{Scene 3 -- ISO 25600.}
        \vspace{0.1cm}
   	\end{subfigure}
   	\begin{subfigure}[t]{\textwidth}
   		\setlength{\tabcolsep}{0.75pt}
   		\centering
   		\begin{tabular}[b]{rccccccc}
   			\includegraphics[height=0.23\textwidth]{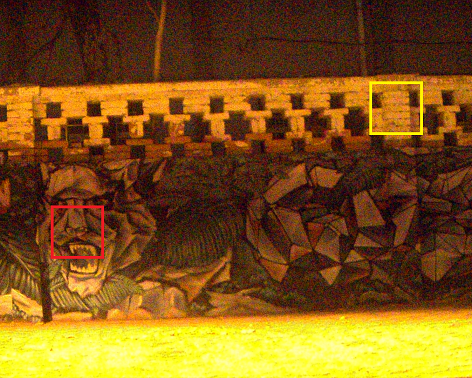} & 
   			\begin{tabular}[b]{c}
   				\includegraphics[width=0.112\textwidth]{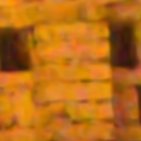} \\[-0.025cm]
   				\includegraphics[width=0.112\textwidth]{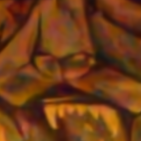}
   			\end{tabular}&
   			\begin{tabular}[b]{c}
   				\includegraphics[width=0.112\textwidth]{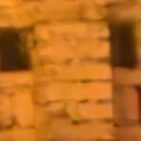} \\[-0.025cm]
   				\includegraphics[width=0.112\textwidth]{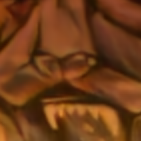}
   			\end{tabular} &
   			\begin{tabular}[b]{c}
   				\includegraphics[width=0.112\textwidth]{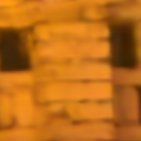} \\[-0.025cm]
   				\includegraphics[width=0.112\textwidth]{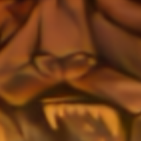}
   			\end{tabular} &
   			\begin{tabular}[b]{c}
   				\includegraphics[width=0.112\textwidth]{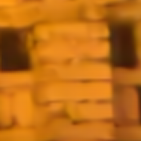} \\[-0.025cm]
   				\includegraphics[width=0.112\textwidth]{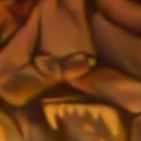}
   			\end{tabular} &
   			\begin{tabular}[b]{c}
   				\includegraphics[width=0.112\textwidth]{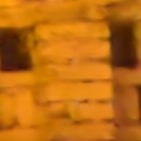} \\[-0.025cm]
   				\includegraphics[width=0.112\textwidth]{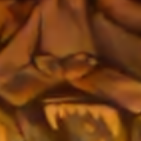}
   			\end{tabular} &
   			\begin{tabular}[b]{c}
   				\includegraphics[width=0.112\textwidth]{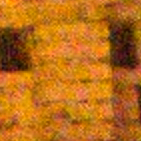} \\[-0.025cm]
   				\includegraphics[width=0.112\textwidth]{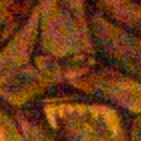}
   			\end{tabular}
   		\end{tabular}
        \vspace{-0.55cm}
   		\caption{Scene 9 -- ISO 25600.}
   	\end{subfigure}
    \caption{Visual comparisons on SONY IMX385 1080p videos from the CRVD dataset~\cite{yue2020supervised}. The proposed EMVD exhibits more details and better noise suppression than more complex state-of-the-art methods in both static and dynamic regions.}
    \label{fig:visual_results} 
    \vspace{-0.1cm}
    \end{figure*}

    \begin{table}[t!]
        \centering
        \small
        \begin{tabular}{ccccc}
            \hline  

            Model                     & GFLOPs      & raw                               & sRGB                               \\
            \hline\hline
            EDVR                      & 3088.98     &\textbf{44.71} / \textbf{0.9902 }  & \textbf{40.89} / \textbf{0.9838}   \\
 
            RViDeNet                  & 1965.04     & 44.08 / 0.9881                    & 40.03 / 0.9802                     \\
            FastDVDnet                & 664.99      & 44.30 / 0.9891                    & 39.91 / 0.9812                     \\
            Ours                      & 79.52       & 44.05 / 0.9890                    & 39.53 / 0.9796                     \\
            \hline

            FastDVDnet$^\dagger$      & 22.16       & 42.25 / 0.9806                    & 37.43 / 0.9693                     \\
            
            \rowcolor{LightYellow}
            Ours                      & 5.38        & 42.63 / 0.9851                    & 38.27 / 0.9722                     \\
            VBM4D                     & -           & -                                 & 35.20 / 0.9577                     \\
            \hline 
        \end{tabular}
        \caption{Objective performance on the CRVD dataset~\cite{yue2020supervised}. Reduced complexity is denoted by $^\dagger$.}
        \label{tab:CRVD}
    \end{table}
    
    \begin{figure}[ht!]
        \centering
        \includegraphics[trim=0.25cm 0.2cm 0.2cm 0.25cm,clip,width=0.99\columnwidth]{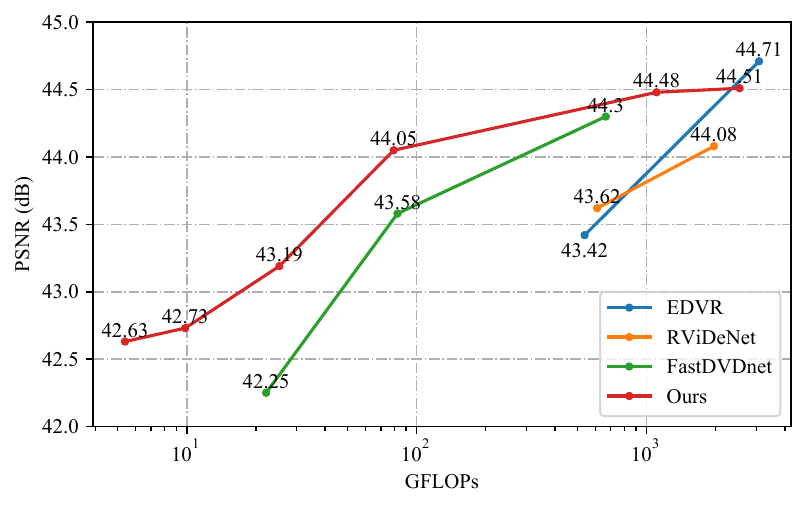}
        \vspace{-0.25cm}
        \caption{Performance (PSNR) of different models at various complexity levels (GFLOPs) on the raw CRVD dataset~\cite{yue2020supervised}.}
        \vspace{-0.15cm}
        \label{fig:psnr_gflop} 
    \end{figure}

    \subsection{Results} 
    
\begin{figure}[t!]
        \centering
        \includegraphics[trim=0.28cm 0.15cm 0.2cm 0.15cm,clip,width=0.99\columnwidth]{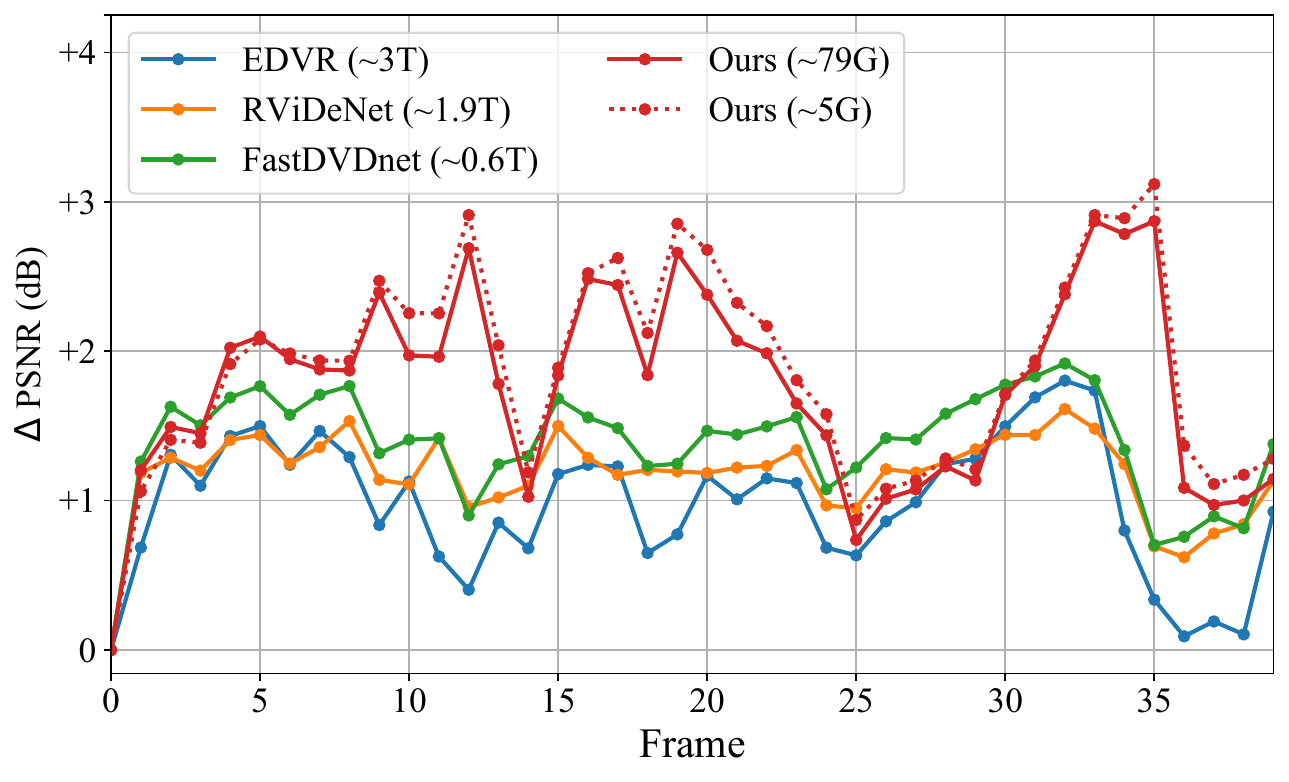}
        \vspace{-0.1cm}
        \caption{Frame-by-frame increase of PSNR with respect to the first frame of the sequence using the SRVD dataset~\cite{chen2018sid,yue2020supervised}. Complexity (FLOPs) is reported in parenthesis.}
        \label{fig:psnr_frames}
        \vspace{0.2cm}
    \end{figure}

    Table~\ref{tab:CRVD} and Fig.~\ref{fig:psnr_gflop} show objective results for all compared models. We denote with $\dagger$ low-complexity implementations which we obtain by evenly reducing the number of convolutional layers and channels. Our experiments indicate that the proposed EMVD significantly outperforms all compared methods when GFLOPs is lower than 100, and the improvement in PSNR increases to more than 1dB as the complexity decreases. Note that we do not provide results for RViDeNet and EDVR with complexity lower than 100 GFLOPs because such models fail to converge in that range. Details of the network structures of the proposed method at varying level of complexity can be found in Table~\ref{tab:distribution}. More technical implementation details are discussed in the supplementary materials. \\
    \indent Interestingly, EMVD is even able to maintain a good margin over state-of-the-art methods with significantly higher complexity. For instance, if we compare the \textapprox79~GFLOPs EMVD against a 25$\times$ larger RViDeNet we observe only a 0.03dB loss in PSNR. This confirms the ability of our method to operate under stringent computational budgets without significant loss in performance and image quality. In Fig.~\ref{fig:visual_results} we show the visual comparison for several test videos in the CRVD dataset captured at ISO 25600. The bricks and trees in the background indicate that EDVR and FastDVDnet generate better details than RViDeNet. The proposed EMVD generates the most pleasing visual results in both static and dynamic regions despite having a complexity 573$\times$ lower than EDVR and 364$\times$ lower than RViDeNet.\\
    \indent In Fig.~\ref{fig:psnr_frames} we analyze the temporal behavior of the compared models using the SRVD dataset (MOT17-01 synthesized ISO 25600)~\cite{chen2018sid,yue2020supervised}. In particular, the plot shows the frame-by-frame PSNR difference ($\Delta$) with respect to the initial frame of the sequence. We note that the $\Delta$ PSNR of RViDeNet, FastDVDnet, and EDVR is --on average-- stable after an initial warm-up period. As a matter of fact, performance of multi-frame methods is bounded by the amount of frames that these models are designed to process at any given time. Differently, the proposed EMVD is a recurrent method, thus by accumulating long-term temporal dependencies it is able to achieve a significantly higher $\Delta$ PSNR in the majority of the sequence. However, multi-frame methods have a slight advantage in dynamic scenes (i.e., frame 25--30) because they can access future frames. Nevertheless, even in these cases the proposed EMVD is able to recover very quickly, as it outperforms all other methods by more than 2dB PSNR after processing only a few frames.\\
    \indent Finally, Table~\ref{tab:soc_profiling} reports on-chip running time and memory usage profiled on a commercial SoC (Huawei P40 Pro Smartphone) using the AI benchmark tool~\cite{ignatov2019ai}. Results demonstrate that our method can attain real-time performance (\textapprox30fps) while still outperforming the reference low-complexity method FastDVDNet in terms of computational requirements (4.9$\times$ faster inference, 6.5$\times$ less memory) as well as objective performance (0.84dB better PSNR).

    \begin{table}[t]
        \centering
        \small
        \setlength{\tabcolsep}{1.75pt}
        \begin{tabular}{cccccc}
            \hline
            Model                & GFLOPs & Time (ms)& DDR (MB) & PSNR / SSIM                      \\
            \hline \hline
            FastDVDnet$^\dagger$ & 22.16  & 177      & 724      & 37.43 / 0.9693                   \\        
            \rowcolor{LightYellow}
            Ours                 & 5.38   & 36       & 112      & \textbf{38.27} / \textbf{0.9722} \\
         \hline
        \end{tabular}
        \caption{Running time and DDR memory required to process a single-precision 720p sequence on a Huawei P40 Pro. Models have been profiled with the AI benchmark tool~\cite{ignatov2019ai}.}
        \label{tab:soc_profiling}
    \end{table}

    \section{Conclusions}
    \label{section:conclusion}

    In this work we have proposed EMVD, an efficient video denoising method which recursively exploit the spatio-temporal correlation inherently present in natural videos through multiple cascading processing stages applied in a recurrent fashion, namely temporal fusion, spatial denoising, and spatio-temporal refinement.

    This multi-stage design, coupled with learnable and invertible decorrelating transforms, allows to significantly reduce the model complexity without seriously impacting its performance. It is interesting to note that the CNNs employed in each individual stage converge to the desired behavior without explicit supervision (i.e., extra terms in the loss), hence making the proposed model straightforward to train. Further, we can gain insights on the inner workings of the model by inspecting the output at every stage (Fig.~\ref{fig:fusion} and Fig.~\ref{fig:refinement}) which in turn allows to interpret and disentangle the effects of spatial and temporal processing.
    
    The proposed EMVD 1) significantly outperforms other state-of-the-art video restoration methods when complexity is constrained, and 2) even remains competitive when complexity of compared models is several orders of magnitude higher (Fig.~\ref{fig:psnr_gflop}). Further, we have verified that EMVD achieves real-time performance (\textapprox30fps at 720p) on a commercial SoC (Table~\ref{tab:soc_profiling}) with a limited memory footprint, thus demonstrating that the proposed method can be practically employed for video processing on mobile devices.

    {\small
        \bibliographystyle{ieee_fullname}
        \bibliography{biblio}
    }
    
\end{document}